\documentclass{article}

\usepackage{arxiv}

\usepackage[utf8]{inputenc} 
\usepackage[T1]{fontenc}    
\usepackage{hyperref}       
\usepackage{url}            
\usepackage{booktabs}       
\usepackage{amsfonts}       
\usepackage{nicefrac}       
\usepackage{microtype}      
\usepackage{graphicx}
\usepackage{natbib}
\usepackage{doi}
\usepackage{amsmath}
\usepackage{cleveref}       

\title{Errorless Irrationality: A unified computational account of the inverse base-rate effect across predictive, observational, and unsupervised procedures}

\usepackage{authblk}

\newbox{\orcid}\sbox{\orcid}{\includegraphics[scale=0.06]{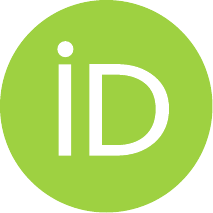}}
\author[1,2]{%
	\href{https://orcid.org/0000-0001-7487-1974}{\usebox{\orcid}\hspace{1mm}Lenard~Dome\thanks{\texttt{lenard.dome@uni-tuebingen.de}.}}%
}
\author[3]{%
	\href{https://orcid.org/0000-0003-4803-0367}{\usebox{\orcid}\hspace{1mm}Andy~J.~Wills\thanks{\texttt{andy.wills@plymouth.ac.uk}}}%
}
\affil[1]{Department of Psychiatry and Psychotherapy, Faculty of Medicine, University of Tübingen, Tübingen, DE}
\affil[2]{German Center for Mental Health (DZPG), Tübingen, DE}
\affil[3]{School of Psychology, Faculty of Health, University of Plymouth, Plymouth, UK}

\renewcommand{\shorttitle}{Errorless Irrationality}

\hypersetup{
pdftitle={Errorless Irrationality: A unified computational account of the inverse base-rate effect across predictive, observational, and unsupervised procedures},
pdfsubject={q-bio.NC, q-bio.QM},
pdfauthor={Lenard Dome, Andy J.~Wills},
pdfkeywords={auto-associator network, neural network, irrationality, computational model, inverse base-rate effect},
}

\begin{document}
\maketitle

\begin{abstract}
	The inverse base-rate effect is a robust bias in how people resolve ambiguity between competing categories, and the most prominent theories explain it through prediction error. Across two experiments we progressively removed the elements of the predictive-learning design that supply such error signals: first by moving to observational learning, then to an unsupervised procedure in which category labels were not presented. The effect persisted--the irrational bias is independent of supervised learning procedures. We propose a new theory, OSCAR, that integrates core computational principles of the best-validated models and operates on self-generated feedback akin to pattern completion. OSCAR extends the learning dynamics underlying the response bias to observational and unsupervised procedures. Evaluated on a large preexisting supervised dataset in addition to the two new experiments reported here, OSCAR performs competitively against alternatives, and is the first model that reproduces the pattern of individual differences seen in humans across all three procedures. The model provides an explanation of hitherto unexplained eye-tracking data, something none of the alternative accounts  provide.

\end{abstract}


\section{Introduction}


The \textit{inverse base-rate effect} \citep[IBRE,][]{medin1988problem} is an irrational tendency in humans to overweigh rare events when faced with ambiguity. In a traditional design, people learn to categorise two overlapping sets of features under two distinct category labels. These sets share a single feature, $A$, and possess a unique feature, $B$ and $C$, predictive of their respective category label. The training thus can be summarised under two trial types, which we will express as $AB \to common$ and $AC \to rare$. During learning, these sets of features occur at different frequencies. The features under the common label usually occur three times as often as features under the rare label \citep{kruschke1996base}. Following training, people categorise features presented by themselves and in novel combinations. People tend to optimally label uniquely predictive features, $B$ and $C$, with their respective common and rare labels when presented by themselves. Responses on the shared feature $A$ tend to show the base-rate following, $A \to common$. But when uniquely predictive features are paired, $BC$, people tend to respond with the rare category label. According to Classical Probability Theory, the rational response is to categorize this ambiguous combination under the common label, because it is the most frequently occurring label. This rare bias on ambiguous combinations of BC has been observed across a variety of experimental manipulations \citep{kalish2001inverse,don2017effects,don2021attention,inkster2022effect,wills2014attention}. For a more thorough introduction to this irrational bias, see a review by \citet{don2021hearing}.

\subsection{Theories of the IBRE}

The most prominent theories of the IBRE involve an attentional mechanism that drives both learning and responding \citep{kruschke1996base, kruschke2001toward}, and whose theoretical roots date back to \cite{mackintosh1975theory}. These explanations rely on a process that relocates attention in response to prediction errors - they update attentional values according to gradient descent. Their explanation is simple. During learning, people learn to label the $AB$ compound first, but they are still learning to label the $AC$ compound. The presence of $A$ tends to push participants to generalize what they learned about $AB$, so they label $AC$ as common, which results in an error. After making this error, attention relocates towards the uniquely predictive feature $C$ to reduce future errors. This results in $C$ acquiring higher attentional salience than $B$. When the ambiguous $BC$ compound is presented, this attentional allocation persists and thus $C$ will dominate responding. This results in an irrational tendency to respond with the rare label. According to these models, this irrationality results from an optimisation process that tries to reduce the errors people make. This process creates an asymmetric cognitive representation that can be summarized as $AB$ belongs to common, $AB \to common$, and C belongs to rare, $C \to rare$ \citep{kruschke2001inverse}. 

\subsection{Current Study}

In this work, we intend to test this basic assumption of the attentional explanation. In the following two experiments, we will gradually remove components from the design traditionally associated with prediction error. Our overarching goal is to investigate whether we can still observe the IBRE, even if we experimentally remove a crucial assumption of existing accounts. In our first attempt, we implement the canonical IBRE design with a caveat that category labels are presented in unison with features.

In our second attempt, we further remove the causal framing of the relationship between features and category labels. The goal was to remove any design component that might affect attentional allocation or the development of asymmetric representation in response to errors. Any presumption of a causal relationship might inadvertently relocate attention in line with the direction of causality between features and labels.

\subsection{Related Work}

To our knowledge, there is only one attempt to implement the standard trial-by-trial IBRE procedure without explicit feedback. In terms of a clear observational-learning version of the IBRE, \cite{johansen2007paradoxical} included the result of a short pilot experiment in their Appendix. Unfortunately, there is no statistical analysis confirming that the IBRE is reliably observed. \cite{johansen2007paradoxical} report a sample size of 16. If we use an effect size of $d = 0.46$ observed by \cite{inkster2022neural} and an $\alpha$ of 0.05 with a non-directional alternative hypothesis, the experiment has 24\% power\footnote{We used the method provided by the R package pwr \citep{champely2020pwr} to calculate power.}. Given this information, this pilot experiment is underpowered. There are also no details about the procedure of this experiment. Therefore, we cannot make direct comparisons.

Nonetheless, \citet{johansen2007paradoxical} demonstrated that the inverse base-rate effect can occur without the traditional predictive learning design. In one of the conditions in their Experiment 3, the canonical inverse base-rate design (including the shared cue) was implemented in a list format. In this format, the trial-by-trial presentation of training items was turned into a list of 12 items fitted on a single page. Subsequently, participants made judgements about new cases on a separate page. In this condition, participants still exhibited a rare preference on $BC$ trials. In another condition of Experiment 3, participants received the information about outcome frequencies as a summary before testing. This summary was presented as prose. After learning about feature-label information in this manner, participants did not show the IBRE but was matching the base rate. 

Additionally, there are at least three studies which are taken as evidence for attentional-reallocation processes in the IBRE. In an eye-tracking study, \citet{don2019learned} demonstrated that on $AC$ trials, people fixated on $C$ longer than on $A$ both pre-responding during stimulus presentation and post-responding during feedback  \citep[see also][]{kruschke2005eye}. This fixation bias increased with more training. They also observe greater fixation on $C$ on $AC$ trials, relative to $B$ on $AB$ trials. Further, \citet{wills2014attention} in an EEG study observed posterior selection negativity and concurrent frontal positivity for C relative to B, which provided evidence for attentional reallocation. Both of these studies argue that these results are consistent with models of error-driven attentional reallocation. \citet{inkster2022neural} investigated the same hypothesis with fMRI. Their region of interest (ROI) analysis explicitly targeted areas that were hypothesized to be involved in the computation of prediction error. They showed that these areas exhibited greater activation during the test phase for $C$ relative to $B$ by themselves. Similarly to eye-tracking, differences in brain activations are taken to correspond to the mechanism specified in attentional theories. Given these findings, it is reasonable to suggest that prediction-error-driven attentional reallocation occurs in a standard supervised learning paradigm.

\section{Experiment 1}

Below, we detail our first attempt to test whether we could observe the rare response bias to $BC$ without an explicit error-driven psychological mechanism. The design component which is most likely to result in any error-driven tuning is feedback. To remove feedback, Experiment 1 will present category labels simultaneously with their respective features. We retain the sequential property of the experiment, which means that participants learn about feature and category relationships on a trial-by-trial basis.

\begin{figure}[!hb]
    \centering
    \includegraphics[width=0.75\linewidth]{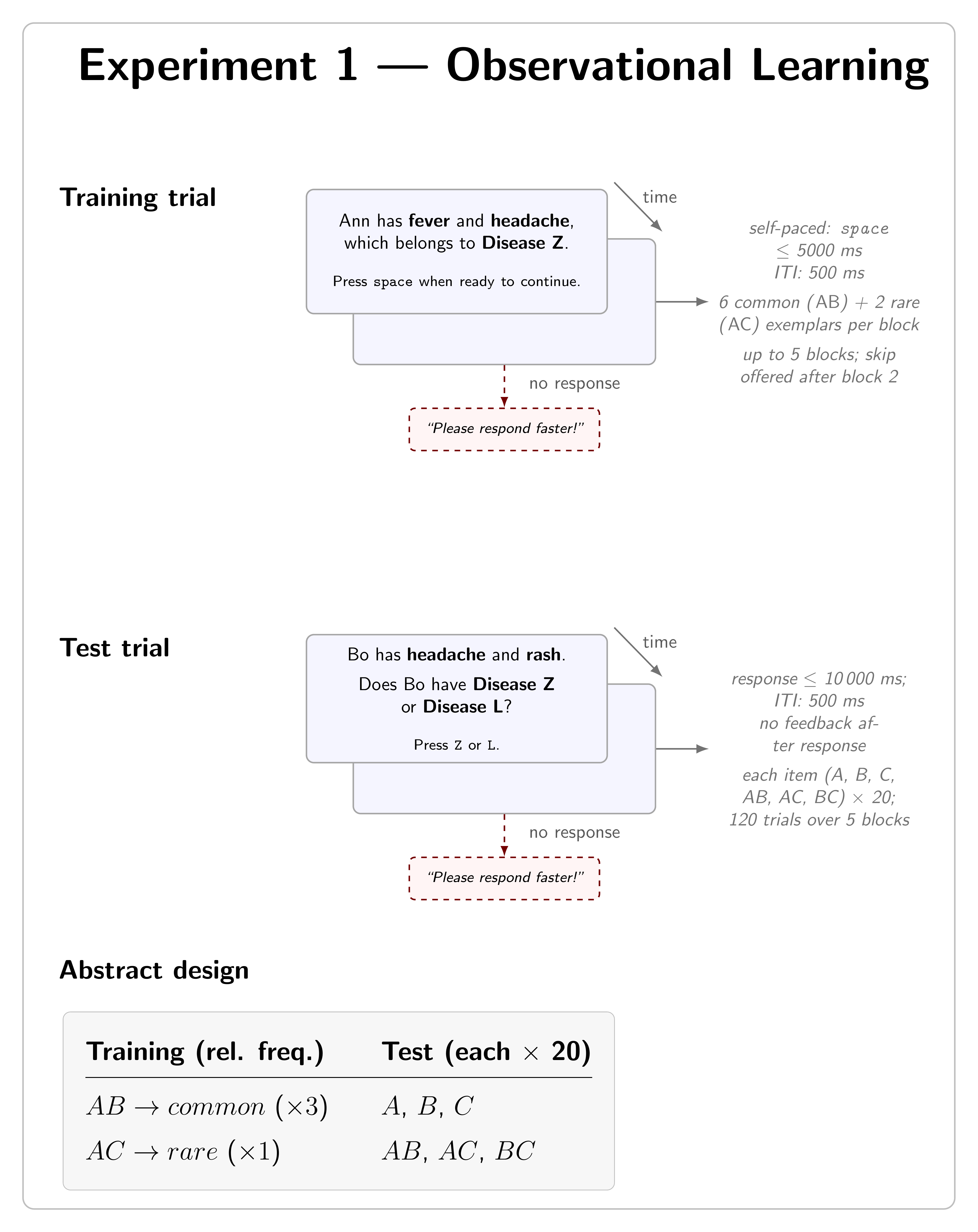}
    \caption{\textbf{Trial structure and abstract design of Experiment 1 (observational learning)}. Symptoms and disease labels appeared together during training and symptoms appeared alone at test. Stacked screens depict the sequence of self-paced trials (diagonal arrows mark the passage of time); dashed branches show the warning displayed when the response deadline passed. The table summarizes the abstract design. Full procedural details are given in the Method sections. \label{fig:experimental-design-a}}
\end{figure}

\subsection{Method}

\subsubsection{Participants}

Participants were undergraduate students who received course credit for their participation. We recruited 169 participants online through the SONA recruitment system.

\subsubsection{Apparatus}

The experiment was programmed in jsPsych \citep{deleeuw2015JsPsych} to be run in a web browser. Participants completed the experiment on their personal computers. The experiment did not allow the use of tablets and smartphones.

\subsubsection{Stimuli}

Category labels corresponded with response keys and were called Disease \textbf{Z} and Disease \textbf{L}. Category features were symptoms: fever, headache, and rash. These physical features were randomly allocated to abstract features, A, B, and C at the beginning of each session. Features and labels appeared in full sentences, such as '\textit{John has fever and rash, which belongs to disease Z}'. Names were randomly drawn from a pool of male and female first names. The list was compiled from an online repository of popular baby names\footnote{The list was taken and later curated from a GitHub repository: \href{https://github.com/aruljohn/popular-baby-names}{https://github.com/aruljohn/popular-baby-names}.}. We selected the 50 most popular male and female names from 2021. Disease names corresponded to response keys and were randomly allocated to either the common or rare category label at the beginning of each session.

\subsubsection{Procedure}

Figure \ref{fig:experimental-design-a} summarizes the abstract design of the experiment. This design is the simplest implementation of the IBRE procedure to date. Participants completed two phases: a training and a test phase. In the training phase, they encountered descriptions of people, the symptoms they experienced, and their respective diseases. These descriptions appeared in the format of '\textit{John has fever and rash, which belongs to disease Z}'. Participants studied these examples and when they were ready to move on, they pressed the spacebar. They needed to complete reading the description within 5 seconds. If the 5 seconds threshold was passed, a screen appeared with the message '\textit{Please respond faster!}'. In each training block, participants encountered 6 common diseases (common category exemplars) and 2 rare diseases (rare category exemplars). After the second block of training, participants were given a choice. They could either move straight to the test phase or complete another training block. A prompt appeared saying that '\textit{Now you have the option to skip the rest of the training phase and move straight to the test phase. If you think you need some more time, you can continue training and study more patients.}'. There were a maximum of 5 blocks they could complete.

In the test phase, participants judged individual symptoms and novel combinations of old symptoms, see Figure \ref{fig:experimental-design-a}. Symptoms appeared in a sentence, such as '\textit{John has a fever.}', with a prompt asking participants to say what disease the person has, '\textit{Does the patient have disease Z or disease L?}'. Participants had to respond by pressing either Z or L on the keyboard. They had 10 seconds to do so, otherwise, a '\textit{Please respond faster!}' message appeared. After the button press, there was no feedback. Each unique test item and training item (occurring in the test phase) was repeated 20 times. So, the test phase included 120 trials, which were broken down into 5 blocks of 24 trials.

\subsubsection{Analysis}

In order to test for the presence of the IBRE, we calculated a Bayes Factor for a one-sample design. We calculate the probability of responding with the rare label on the critical BC test item, $P(rare|BC)$, for each participant. Then we tested this distribution of probabilities against the null, $mu = 0.5$, which denoted random responding. If the Bayes Factor fell below 1/3, we concluded that participants' responses are not different from random responding. If the Bayes Factor fell above 3, we concluded that participants' responses reliably differ from null. If the mean probability of $P(rare|BC)$ is higher than 0.5, we conclude that we observed the IBRE. Values lower than 0.5 would indicate base-rate following. We used the method implemented in the BayesFactor R package \citep{morey2022bayes}.

\subsubsection{Exclusion}

To match performance with the predictive learning implementations of the IBRE, we decided to exclude participants whose test performance on the training items fell below 0.75 accuracy. This level of accuracy was the lowest at which the evidence that the participant performed better than chance was above the Bayes Factor of 3. We calculated the Bayes Factor for binomial proportions via the method implemented in BayesFactor R package \citep{morey2022bayes}.

\subsection{Results and Discussion}

After exclusion, 125 participants made it into our main analysis. In summary, the qualitative pattern in our results corresponds to the base result of the IBRE. Participants exhibited a reliable common preference for $A$, $M_{A} = 0.68$, 95\% HDI $[0.63, 0.73]$, $\mathrm{BF}_{10} = 2.45 \times 10^{7}$. For this cue, people explicitly followed the base rate - responded rationally according to Probability Theory. In contrast, participants showed a reliable rare preference for $BC$, $M_{BC} = 0.67$, 95\% HDI $[0.62, 0.72]$, $\mathrm{BF}_{10} = 1.11 \times 10^{7}$. This gives us a sufficient amount of evidence to conclude that we have observed the IBRE.

Thus the current study strongly confirms that the IBRE can be observed in an observational procedure. In the current experimental design, the IBRE emerged in the absence of an explicit prediction error that drives the development of attentional allocation. All attentional theories of the IBRE rely on the assumption that this irrational rare preference arises as a result of optimising accuracy during the training phase. In the absence of this explicit prediction error, EXIT-like theories cannot predict the presence of the IBRE.

One aspect of the current design is that participants might still experience internally-generated prediction errors from feature to categories on a trial-by-trial basis. Given that the general assumption is that diseases cause symptoms, participants could likely assume a causal link between symptoms and diseases. This assumed causal relationship can encourage participants to make not an explicit but a silent prediction. Informally, participants might think of a certain feature--label causal relationship while reading the sentences. People then resolve errors between the expected and the observed feature--label causality by allocating attention to rare features to distinguish diseases.

In Experiment 2, we adress this by removing any design component that makes it clear to participants what the category label is. We also used stimuli that reduced the chance of people assuming a causal relationship between the stimulus's features.

\section{Experiment 2}

In this experiment, we implemented the IBRE in a procedure similar to a cued-recall task. All stimuli were solid black geometric shapes. The task asked participants to memorize the shapes. On each trial, we randomized the position of the geometric shapes in the arrangement. This further minimized the chances of having any design component suggestive of which feature is the category label.

\begin{figure}
    \centering
    \includegraphics[width=0.75\linewidth]{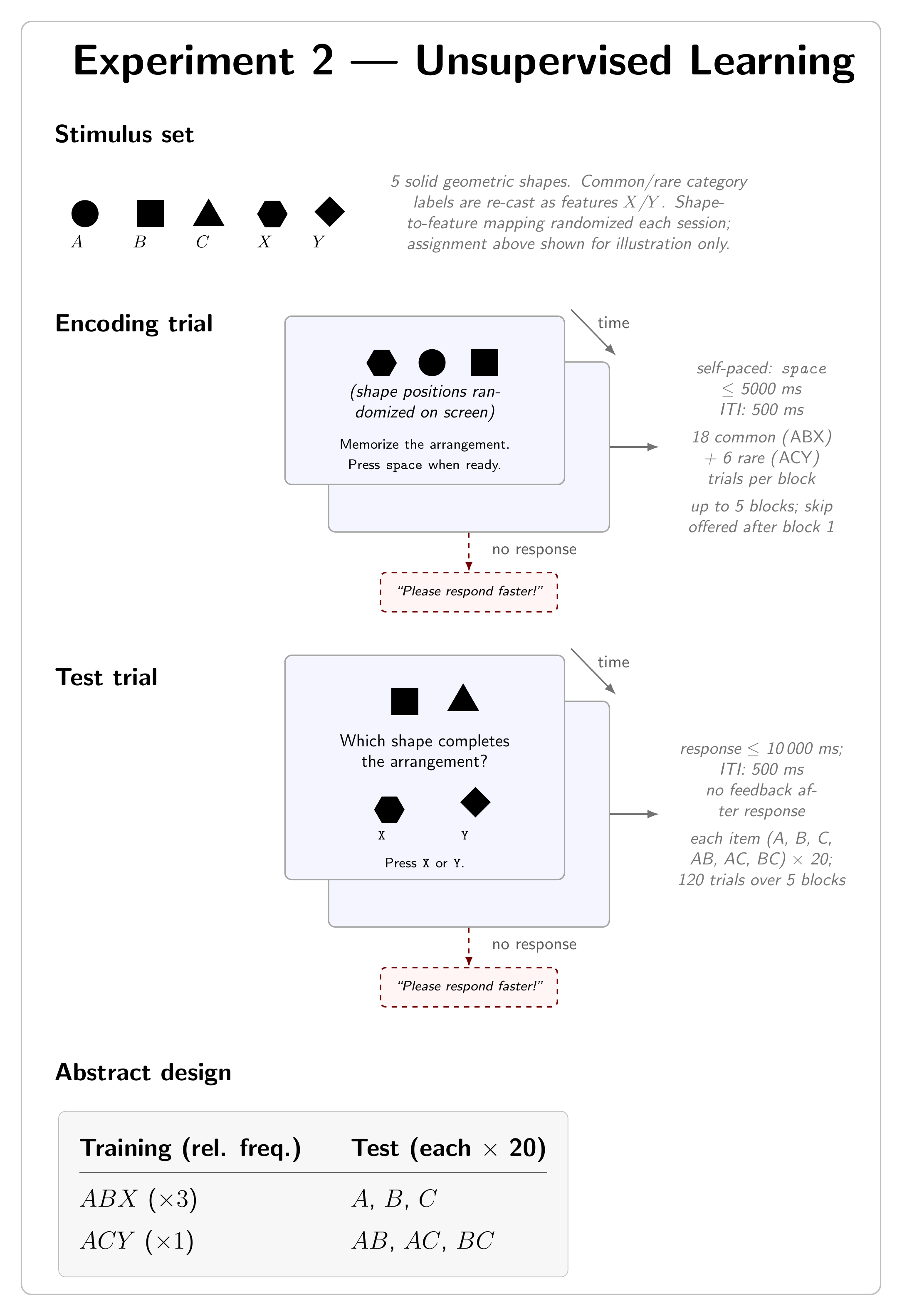}
    \caption{\textbf{Trial structure and abstract design of Experiment 2 (unsupervised learning).} Geometric shapes were randomly mapped to the abstract features A, B, C, X, and Y at the start of each session (assignment shown for illustration only) and the former category labels were re-cast as features X and Y. Stacked screens depict the sequence of self-paced trials (diagonal arrows mark the passage of time); dashed branches show the warning displayed when the response deadline passed. The table summarizes the abstract design. Full procedural details are given in the Method section.}
    \label{fig:experimental-design-b}
\end{figure}

\subsection{Method}

\subsubsection{Participants}

We recruited 171 undergraduate students who completed the experiment for partial course credit. Recruitment was done via the SONA recruitment system. 

\subsubsection{Stimuli}

Stimuli were common solid geometric shapes, shown in Figure \ref{fig:experimental-design-b}. Common and rare category labels were turned into features X and Y respectively. Each shape was randomly allocated to one of the abstract features shown in Figure \ref{fig:experimental-design-b}.

\subsubsection{Procedure}

Figure \ref{fig:experimental-design-b} depicts the abstract experiment design. Similar to the previous experiment, participants completed two phases: an encoding/training and a test phase. In the training/encoding phase, participants were repeatedly exposed to the exemplars and were asked to memorize the arrangement of geometric shapes. Unlike Experiment 1, exemplars were composed of three geometric shapes. On each trial, geometric shapes appeared in different orders, such that the position of features on the screen was completely counterbalanced. This resulted in 24 trials within each block, which contained 18 common trials and 6 rare trials. Similar to Experiment 1, participants could complete a maximum of 5 blocks. Beginning at block 2, they were given a chance after completing each block to move straight to the test phase. The trial structure and response deadlines were the same as Experiment 1.

In the test phase, participants were shown \textit{incomplete} arrangements of geometric shapes and were asked to complete them by selecting either \textbf{X} or \textbf{Y} corresponding to different shapes. Similar to Experiment 1, each test item (incomplete arrangement of shapes) appeared 20 times. The test phase was composed of 120 trials presented across 5 blocks of 24 trials.

\subsubsection{Analysis and Exclusion}

We applied the same analysis and exclusion methods as in Experiment 1.

\subsection{Results and Discussion}

After exclusion, 86 participants made it into our analysis. The results are a qualitative and ordinal match to Experiment 1. Participants showed a clear common preference for stimuli A, $M_{A} = 0.78$, 95\% HDI $[0.73, 0.83]$, $\mathrm{BF}_{10} = 5.37 \times 10^{13}$. Participants also showed a reliable rare preference on ambiguous BC trials, $M_{BC} = 0.73$, 95\% HDI $[0.67, 0.79]$, $\mathrm{BF}_{10} = 8.12 \times 10^{8}$. This gives us a sufficient amount of evidence to conclude that we have observed the IBRE. Here, we further demonstrated that the IBRE can arise without experimental-design components that explicitly promote an error-driven process.

\section{Modelling}

Across two experiments, we tested a central assumption of the most prominent theories of the IBRE. This central assumption was that the IBRE is caused by the presence of prediction error. Together, the experiments establish that IBRE persists in the absence of all design components on which error-driven theories depend. As a consequence, any model that aims to account for the irrational response bias across these experiments as a single cognitive phenomenon -- rather than as a disparate set of results -- must satisfy three requirements:

\begin{enumerate}
    \item \emph{Operate without explicit feedback.} The learning signal must remain available when category labels are presented in unison with features rather than outcomes to be predicted; and in the absence of explicit error-correction signal (Experiment 1).
    \item \emph{Operate without feature-to-label causality.} The learning signal must remain available when the experimental procedure provides no causal cues as to which dimension is the category label (Experiment 2).
    \item \emph{Use a single mechanism across the three different procedures.} Attention, prediction, and learning must be driven by a common signal across the three disparate procedures. While this is not a strict theoretical requirement, a unified account is preferred on grounds of parsimony.
\end{enumerate}

These requirements have strict implications for existing accounts. The issue lies at the level at which these theories are formalized: they offer no mechanism for how learning proceeds without an explicit teaching signal. The commitment to an external label, and hence an externally provided error signal, makes the non-supervised cases hard for feed-forward network models \citep[such as][]{kruschke2001toward}. Alternative models without prediction error, such as the Dissimilarity Generalized Context Model \citep[DGCM][]{stewart2007dissimilarity,o2018model}, sidestep the learning mechanism entirely but as a result, they take the problem outside of the model by providing no learning mechanism.

In what follows, we introduce OSCAR -- an auto-associative network architecture that meets the three requirements in the list above, driving attention, prediction, and learning from a single error signal generated internally based on principles of pattern completion.

\subsection{OSCAR}

To unify attention in learning across supervised, observational and unsupervised paradigms, we require a single mechanism that operates across all three instantiations of the inverse base-rate effect. Auto-associative networks are driven by pattern completion where error is generated across all input dimensions, so they are a natural candidate for the implementation framework because the same signal determines attention, prediction, and learning regardless of the paradigm. In our current formulation, we will assume a feed-forward auto-associative network, which is more similar to modern engineering applications \citep{kramer1991nonlinear,bourlard1988autoassociation,bourlard2022autoencoders}, as opposed to the fully recurrent-network formulation that may be more familiar to those with a background in 1980s connectionist theory in psychology, for example \cite{rumelhart1986parallel} or  \cite{hinton1987learning}.

Based on recent model evaluations \citep{paskewitz2020dissecting, dome2025better, dome2025gdistance}, the most successful instantiation within the error-driven learning class of theories is a Neural Network with Rapid Attention Shift \citep[NNRAS;][]{paskewitz2020dissecting}. Thus, we incorporate its core computational principles, including error-driven attention updating in particular, but re-embed them in an auto-associative architecture, which generates its own error signal. Auto-associators learn to reproduce the complete input pattern (stimulus vector, explained below) in the output layer, such that the input is acting both as a teaching signal and the pattern to be associated. The model's overall goal is pattern completion: whenever partial input is presented to the model, the remainder of a pattern is to be completed.

\begin{figure}[t]
    \centering
    \includegraphics[width=\linewidth]{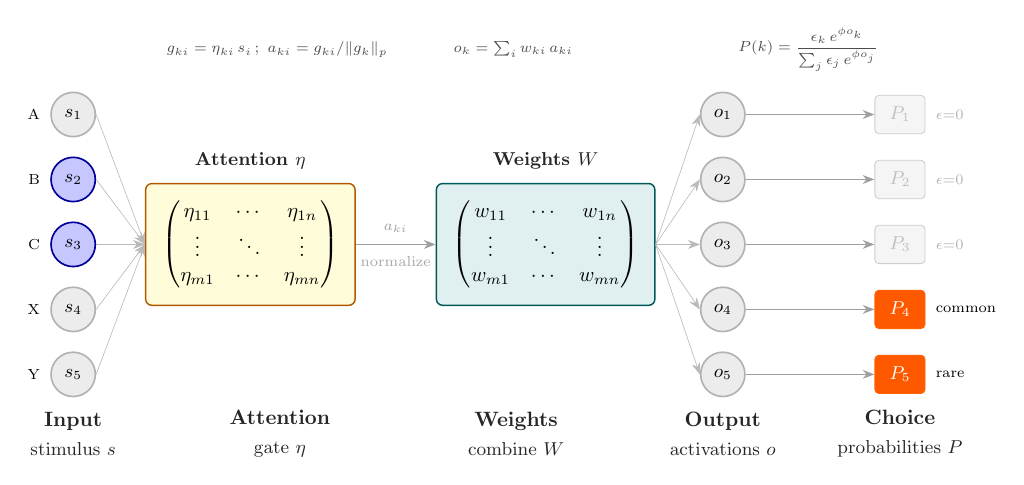}
    \caption{\textbf{Architecture of OSCAR}. A two-layer auto-associator. Input $s$ encodes th stimulus vector, which is combined with the attention hidden layer ($m \times n$ grid; each unit's gain is $g_{ki} = \eta_{ki}s_i$) and is subject to an outcome-specific normalization process resulting in attention strength $a$. Attention is combined with the weight matrix $W$ produce output activations. Choice is then computed via a gated softmax function. Purple nodes show currently active input nodes, $s_i = 1$; and orange nodes show output nodes that are currently excitable, $\epsilon_k = 1$.\label{fig:architecture}}
\end{figure}

In what follows, we formally describe OSCAR (Outcome-Specific Configuration-dependent Attention Representation), a feed-forward auto-associator network with two distinct layers. Figure \ref{fig:architecture} shows its main architecture, and Table \ref{tab:parameters} shows its freely-varying parameters.

\begin{table}[ht]
\caption{List of Model Parameters}
\label{tab:parameters}
        \advance\tabcolsep-1pt
        \centering
        \small
        \begin{tabular}{clc}
          \toprule
          \textbf{Symbols} & \textbf{Description} & \textbf{Range} $[\ell, u]$ \\
          \midrule
          $P$ & p-norm; brutality parameter for attention competition & $[1, 10]$  \\
          $\alpha$ & Weight learning rate & $[0, 1]$ \\
          $\mu$ & Salience learning rate & $[0, 1]$ \\
          $\rho$ & Attention shift rate & $[0, 10]$ \\
          $\phi$ & Response consistency & $[0, 10]$ \\
          $\gamma$ & Irreducible Noise (Lapse Rate) & $[0, 1]$ \\
          \bottomrule
        \end{tabular}
\end{table}

Each input node in the model encodes a distinct stimulus dimension, from herein referred to as features. The model takes a one-dimensional stimulus vector, which is a vector of length equal to $n$ features. This stimulus vector encodes the presence of the $i$th feature as 1 and its absence as 0. The activation of each input is denoted by $s_i$, such that when feature $i$ is present, the input node $i$ activation value is 1, and 0 when its absent. In traditional category learning models, each input node is assumed to have its own feature salience, implemented as dimensional attention strength \citep{kruschke1992alcove, kruschke2001toward}, and represented as a shared global attention vector of length $n$ with each $i$th feature corresponding to the $i$th input node. Saliences are nonnegative and clamped between zero and positive infinity. In prior work, we showed that globally shared attention vectors become unstable in multi-outcome learning and prevent the network from learning to differentially attend to cues \citep{dome2026shared}. An auto-associative network will inevitably encounter this instability because of its nature of predicting multiple active inputs. Following their suggestion, we implemented an attentional weight matrix with dimensional attention vectors in place of a globally shared attention vector, culminating in an attention weight matrix. The dimensional attention vectors allow features to acquire outcome-specific salience, which will selectively activate depending on what the system is trying to predict. This is not unprecedented; there are models of category learning that represented dimensional attention as a vector and not a single point-estimate \citep{kruschke1999model}. In this framework, features can be diagnostic of some outcomes, but not others, which forces attention into a more granular matrix representation, $\eta_{mn}$, where $m = n$. This attention weight matrix sets OSCAR apart from the scalar $\alpha_i$ of the Mackintosh tradition and invites a comparison to self-attention in Transformer architectures \citep{vaswani2017attention}, for an accessible walkthrough, see \cite{raschka2024build}. The comparison should not be pressed. Self-attention allows each position in the input sequence to determine how relevant it is to attend to other positions in the same sequence, which is then incorporated into the representation of said sequence, and updates via backpropagation of error. However, OSCAR learns outcome-specific attention weights to reduce errors, where each attention weight indexes the relevance of the feature to the specific outcome when making a prediction.

OSCAR combines attention weights with the stimulus vector to produce attention gain ($g$):

\begin{equation}
    g_{ki} = \eta_{ki} \times s_i
\end{equation}

where $\eta$ is the underlying salience, representing the tendency of each feature to capture attention. The values in $\eta$ range between 0 to $\infty$. These activations propagate through a competitive gating mechanism that produced an attention matrix ($a$) normalized row-wise: 

\begin{gather}
    a_{ki} = \frac{g_{ki}}{||g_k||_p}\\[10pt]
    ||g_k||_p = (\sum^{n}_{i=1} |g_{ki}|^p)^{\frac{1}{p}}    
\end{gather}

where $||g||_p$ denotes the $p$-norm for the gain vector, $g$. Attention gains are normalized by their vector $p$-norm for the $k$th outcome. This normalization forces features to share a fixed attentional capacity, where the competition for resources are controlled by a brutality parameter, $p$. This parameter determines the degree of competition, with lower $p$ increases competition, and higher $p$ corresponds to less competition. When $p$ approaches infinity, features with the highest gains get the attention of nearly 1. If features are tied for attention gains, they all get attention that approximates 1. When $p = 1$, the attention vector sums to 1, where the increase in attention to one feature will reduce attention to other features.

After the competitive attentional gating, the attentional activations propagate to output units via weighted connections, $W$, which is an $m \times n$ matrix, with $m$ denoting the number of output nodes corresponding to the number of features in the experiment. Here, $w_{ki}$ denotes the connection weight between input $i$ and output node $k$. Output node activations correspond to the activation of category labels:

\begin{equation}
    o_k = \sum_{i}w_{ki}a_{ki}
\end{equation}

The output node activations are then mapped to response probabilities using the \citet{luce1959individual} choice axiom, also known as softmax \citep{bridle1990probabilistic,rumelhart1995backpropagation}. The sensitivity to differences between output  activations is controlled by an inverse temperature parameter, which we denote with $\phi$. Higher values cause the model to become more deterministic, small differences between activations are exaggerated, whereas small values cause the model to become more indecisive.

\begin{equation}
    P(\text{action } k) = \frac{e^{\phi o_k}}{\sum_{j}e^{\phi o_j}}
\end{equation}

In most experimental procedures, the model can only respond with the subset of all possible responses, usually clearly communicated to the participant during the presentation of the stimulus. In most predictive-learning experiments, the model will receive a partial input pattern, and the remainder of the pattern to be completed contains the available choice options. Here, we can naturally assume that all non-present features during the input presentation equal to the available choice options, but it does not need to be the case for all problems. To mediate this, we introduce a binary excitability vector, $\epsilon$, where all nodes corresponding to available choice options will receive an external excitability signal, $\epsilon_j = 1$, and the remaining nodes receive none, $\epsilon_{k \neq j} = 0$. This vector will act as a gating mechanism in the softmax:

\begin{equation}
    P(\text{action } k) = \frac{\epsilon_k e^{\phi o_k}}{\sum_{j}\epsilon_j e^{\phi o_j}}
\end{equation}

After the model predicts a response, it receives feedback in the form of a teaching vector, $t$, with length $m$. 
The teaching vector presents the complete pattern to be learned (both partial input and remainder patterns; $t = \epsilon + s$). In unsupervised cases, $t = s^T$. Error is then calculated between the teaching signal and the output activations prior to gating:

\begin{equation}
    \delta_k = t_k - o_k
\end{equation}

This prediction error is incorporated into a loss function as the sum squared deviation between teacher and output activations, that is derived from stochastic gradient descent:

\begin{equation}
    E_k = \frac{1}{2}\delta_k^2\text{ ,}
\end{equation}

This error-term is similar to the one that has been extensively used to obtain learning rules \citep{rumelhart1986learning,kruschke1992alcove,kruschke2001toward,gershman2017dopamine,paskewitz2020dissecting}, but was prone to produce unstable learning of attentional salience, causing large updates to hit the lower boundary of zero and features to fail to acquire salience \citep{dome2025gdistance}. \cite{dome2026shared} showed that the outcome-specific attentional representation we employ here solves this problem for the following attention updates. Then, attention is then adjusted on gradient descent on error with respect to the underlying gains:

\begin{align}
\Delta g'_{ki} &= \tanh\left(-\rho \frac{\partial E_k}{\partial g_{ki}}\right) \\
&= \tanh\left[\rho s_i \|g'_k\|_p^{-1}\delta'_k(W_{ki} - a'^{p-1}_{ki}o'_k)\right]
\end{align}

where $\rho$ is a positive constant, denoting the step size for the gradient descent, called the attention shift rate; $\tanh$ is a squashing hyperbolic tangent function that we apply to further constrain updates to lie between -1 and 1. Within this framework, attention shifts away from non-informative items towards predictive ones within a single trial. This is a highly nonlinear shift, since the direction and magnitude of the descent is sensitive to the changes in attention: the gradient changes as attention changes \citep{kruschke2001toward}. As a result, the attention shift reiterates ten times, where attention, predictions, and errors recalculated at each step. For each iteration, we further squashed updates via a hyperbolic tangent function, which helped constrain estimates and avoid numerical overflow. This mechanism is also inherited from previous models \citep{paskewitz2020dissecting, kruschke2001toward}. The resulting attention weights are used to update the underlying saliences via displacement, similar to reconstruction error in recirculation networks \cite{hinton1987learning,oreilly1996biologically}, where updates depend on the difference of activations between start and end states of gradient descent: 

\begin{align}
    \Delta \eta_{ki} = \mu (g_{ki} - {g'}_{ki}) s_i
\end{align}

where $\mu$ is the attentional learning rate. The model then uses the updated error and normalised attention matrix to update connection weights according to the delta rule \citep{rescorla1972theory, rumelhart1986learning} with a slight modification to incorporate attention weights \cite{kruschke2001toward}:

\begin{equation}
    \Delta w_{ki} = \alpha \big({\delta'}_k {a'}_{ki} s_i\big)
\end{equation}

where $\alpha$ is a constant learning rate. 

\subsubsection{From weights to eye-tracking}

OSCAR represents attention as a matrix, so moving from a matrix to an ordinal relationship of relative fixation time requires further assumptions that collapse the matrix representation into a point-estimates. Within the \cite{mackintosh1975theory} framework, attention to a feature is determined by its predictiveness: attention is directed towards highly predictive features and away from non-predictive features. Or inversely, uncertainty guides attention: attention is directed away from high uncertainty towards low uncertainty \citep{speekenbrink2022chasing}. We implement this mapping as a context-dependent one-shot process by combining two types of uncertainties. First, we turn the product of attention $a$ and weights $w$ into a distribution via a SoftMax over the prediction space, indexed by $k$, for each $i$th feature separately:

\begin{equation}
    d_{k,i} = \dfrac{e^{a_{k,i} w_{k,i} }}{\sum_{\ell=1}^{K}e^{a_{\ell,i} w_{\ell, i}}}
\end{equation}

Then we can calculate a baseline salience as concentration of certainty on outcome nodes:

\begin{equation}\label{equation:certainty-shannon}
    cert_i = 1 - \frac{H(d_i)}{\ln k}
\end{equation}

where $H(\cdot)$ is the Shannon's Entropy\footnote{$H(X) = -\sum_{x} p(x)\log_2 p(x)$, the average uncertainty of a discrete random variable $X$ in bits.} \citep{shannon1948mathematical}, quantifying the degree of uncertainty of the distribution, and $\ln{k}$ is a normalization constant. Features whose predictions are flat across outcome nodes would acquire a lower values, and features whose prediction over outcome nodes are sharp would acquire higher ones.

Because most stimuli are composed of various features, attentional allocation happens within a feature compound. Features whose predictions share a consensus or disagree equally will have a more uniformly distributed fixation pattern. If one feature has a flat predictive distribution over outcomes is paired with one whose predictions are much sharper, attention should move towards the one with the sharper distribution. In order to quantify that, we will specify a \emph{product-of-experts} that measures the degree of consensus reached with the current feature configurations for each output unit:

\begin{equation}
    c_k = \prod_{i \in S} d_{k, i}
\end{equation}

where agreements are sharpened and disagreements are flattened. We again apply Shannon's entropy to determine the decisiveness of the consensus in a fashion similar to Equation \ref{equation:certainty-shannon}:

\begin{equation}
    dec(S) = 1 - \frac{H(\hat{c})}{\ln{k}}
\end{equation}

where $\hat{c}$ is the normalized consensus, and $dec(S)$ evaluates to a high number when cues agree and a low number when cues do not. Below, we calculate $F$, a net score of fixation:

\begin{equation}
    F_i = \frac{cert_i}{1 + dec(S)}
\end{equation}

The division by $dec(S)$ signals the degree of attentional allocation that needs to take place: when agreement between features is strong, attention is broad and equally distributed between features; but when agreement between features is weak, attention is directed towards the cue with the highest certainty, $cert$. When features are tied in their certainty, but are in disagreement, attentional allocation is also tied between them.

Fixation scores are turned into probabilities of fixation, $F$, for the $i$th feature, given a stimulus $S$ containing $n$ features:

\begin{equation}
    P(i | F_i, S) = \frac{e^{F_i}}{\sum_{j=1}^n e^{F_j}}(1-\gamma) + \frac{\gamma}{n}
\end{equation}

where $(1-\gamma) + \frac{\gamma}{n}$ is called an irreducible noise, quantifying the degree of interference from items other than the fixated. This mixture of softmax policy and uniform distribution over non-focused items originates in \cite{talmi2009integrate} and its current form is first presented by \cite{guitartmasip2012gonogo}. This rule has also been referred to as a lapse rate in psychophysics \citep{wichmann2001psychometric}, and tremble in game theory \citep{selten1975reexamination}. In relative terms, irreducible noise raises the fixation proportions for stimuli whose fixation score is exceptionally low. This may seem counterintuitive, but it can account for noisy fixation lapses. 

\subsection{Model Evaluation Framework}

We evaluated OSCAR across three learning regimes: supervised, observational, and unsupervised. Model comparison was restricted to the supervised dataset of \cite{dome2025gdistance}, the standard protocol under which the competing IBRE models were developed. Those architectures cannot be applied to the observational and unsupervised datasets introduced here: without an explicit teaching signal, they generate no error to learn from. For the supervised dataset, we included a Neural Network with competitive attention gating and Rapid Attention Shift, NNRAS \citep{paskewitz2020dissecting}; a long-time title holder EXemplar-based attention to distinctive InpuT model \citep[EXIT;][]{kruschke2001toward}; and a Dissimilarity Generalization Context Model \citep[DGCM;][]{o2018model,stewart2007dissimilarity}. Their respective parameters and parameter bounds are presented in Supplementary Table \ref{tab:adaptive-bounds}.

We evaluate models on two independent frameworks: the first is $g$-distance \citep{dome2025gdistance}, and the second is conventional goodness-of-fit.  Both evaluations ask whether the model reproduces human behavior, but at different granularities. While goodness-of-fit estimates how well a model numerically approximates a single best-fitting parameter set, $g$-distance asks what can the model produce and how much of that lies within and outside of the set of behaviors humans produce. It estimates this model property through establishing two constructs, which also makes this problem tractable. First, $g$ operates over a discretized result space, and requires \emph{ordinal patterns}, which are discretized summaries of participant behaviors as a set of claims (e.g. participants have preferred A over B over the course of the experiment) and represents the resolution at which the theories under comparison actually make claims. Second, $g$ requires us to enumerate the model's complete \emph{behavioural repertoire}, which is a set of ordinal patterns it can produce anywhere within its bounded parameter space. Below, we first introduce $g$-distance, then the discretization and the enumeration procedures. Derivations and justifications are given in \citet{dome2025gdistance}.

Furthermore, we evaluated model performance on individual data. Each participant across the three datasets experienced a unique trial-order. All models are trial-order sensitive, so model performance is conditioned on the specific trial order each participant experienced. This meant that all routines detailed below were applied to each unique trial order, and the values we report here were aggregated across all trial orders.

\subsubsection{$g$-Distance}

$g$-Distance \citep{dome2025gdistance} is a theory-oriented evaluation framework that quantifies the discrepancy between empirical human data and a model-produced set of generated behaviors within a discretized result space. In more technical terms, $g$ is metaheuristic in multi-objective optimization -- a distance to a reference-object method \citep{collette2013multiobjective}; it quantifies the distance between the model under evaluation and hypothesized perfect model under complete information. Given a model's full behavioral repertoire and the empirically observed ordinal patterns, $g$ is decomposed into two easily interpretable components: $\alpha$ (accommodation) indexes whether the empirical ordinal pattern lies within the model's partitions; $\beta$ (breadth of unobserved model patterns) penalizes the model after each ordinal pattern it produces that are not also part of the observed empirical set of patterns. Through these two components, $g$ is sensitive to model inadequacies invisible to goodness-of-fit, such as qualitative model failure \citep{wills2012adequacya} through $\alpha$; architectural flexibility \citep{gregg1967process}, and excess predictions \citep{roberts2000how} through $\beta$. These dimensions can be differentially weighted through $w$, depending on the belief about the relative importance of accommodation and flexibility. Within this bounded space, a model that perfectly corresponds to the human data would have $g = 0$, which is decomposed into perfect accommodation, $\alpha = 1$, and specificity, $\beta = 0$. This is termed the \emph{PAS} point and is used as the standard reference point in calculating $g$.

\subsubsection{Discretization}\label{section:discretization}

The algorithm computing $g$ operates on ordinal patterns rather than continuous model outputs. For the current work, we have adopted the method applied by \cite{dome2025gdistance}, where continuous output was transformed into an inequality matrix, a symmetrical, non-weighted, and directional adjacency matrix. This matrix depicts the participant's choice profile by contrasting choice probabilities of each cue against all other cues. The comparison records three relationship between cue contrasts: approximate equality; greater; and smaller. The boundary between meaningful and null differences set by a resolution parameter, which were estimated via a Bayesian version of a difference-of-proportions test from the empirical data and similarly applied to model-produced probabilities as well.

\subsubsection{Parameter Space Partitioning} 

All models undergo parameter space partitioning \citep{pitt2006global} over their psychologically plausible parameter bounds (see Table \ref{tab:adaptive-bounds}). Briefly, parameter space partitioning looks for disjoint regions in the parameter space that elicit specific discretized patterns of model behavior. At the end, parameter space partitioning results in a countable set of distinct ordinal patterns the model can produce across its complete parameter space; the procedure enumerates the complete behavioral repertoire of the model. The partitioning of the parameter space follows the procedure set out in \citet{dome2025gdistance}, where the sampling scheme and the criteria defining the accommodation and specificity regions are given in full; see also \cite{dome2024psp}.

\subsubsection{Zone of model adequacy} We further benchmark model performance against the minimum expected performance threshold on participant data given the noise ceiling, which we previously referred to as \emph{zone of model adequacy} \citep{dome2025gdistance}. The zone for each procedure is estimated by a split-half human fit: participant data are repeatedly bisected, and $g$-distance between the two halves are computed. The mean across random splits estimates the expected $g$ of human-level ``model'' predicting human data.

\subsubsection{Scope of evaluation} Of the models compared here, DGCM produces predictions only over the test phase behavior; all others additionally produce learning trajectories. We restrict the model evaluation for the test phases and exclude learning phase responses from the discretization process and the computing of goodness-of-fit. This preserves the comparability of the model set and focuses the evaluation on the shared prediction space. It also functions as a structural train-test separation: for each model that has a learning phase, the data used to drive learning are distinct from the data used to evaluate it. Additionally, we restricted ordinal evaluation to unique test items (A, B, C, BC). Increasing the granularity of the ordinal patterns beyond this is in principle possible, but produces a substantial increase in the time required to enumerate model patterns; we did not pursue it. The granularity ceiling here is set by the computational cost of enumerating partition cells: within that ceiling, the resolution at which each contrast is discretized (\nameref{section:discretization}) is fixed by the empirical resolution of the data rather than chosen arbitrarily.

\subsubsection{Log-likelihood fitting}

To verify that the qualitative ranking obtained through $g$-distance is not restricted to ordinal patterns, we additionally computed a set of goodness-of-fit metrics for each model in the supervised dataset. For the full set of metrics we included in the comparison, see Table \ref{table:metrics}.

We optimized models using a differential evolutionary (DE) algorithm \citep{ardia2011differential, mullen2011deoptim} minimizing the summed negative log-likelihood for test items:

\begin{equation*}
-\log L(\theta \mid Y, M) = -\sum_{i=1}^{N} \log \bigg[ p(y_i \mid \theta) \bigg]
\end{equation*}

where $\theta$ denotes the model parameters, $Y$ represents the dataset containing $N$ data points, and $M$ stands for the model. The term $p(y~|~\theta)$ represents the probability of observing the data $y$ given a specific choice of parameters $\theta$. The DE algorithm iterated 1500 times with a local-to-best mutation strategy, meaning each candidate solution is pulled toward the current best solution during mutation. The scaling factor (F = 1.5) is set aggressively above 1, so the algorithm takes large steps through the parameter space, helping it escape local optima at the cost of occasional overshooting. The crossover rate (CR = 0.2) is kept low, meaning only 20\% of a candidate's parameters are replaced at each step — most of the solution is preserved, keeping changes conservative. The top 25\% of solutions (p = 0.25) are eligible as "best" candidates for the mutation step, and the adaptation constant (c = 0.8) controls how quickly the algorithm updates these control parameters across generations. The resulting subject-level parameter estimates were used to simulate model predictions, as shown on Figure \ref{fig:simulations}A.

\subsection{Computational Results}

We evaluated OSCAR across the three learning procedures in turn. We begin with the canonical, supervised IBRE procedure, which establishes that the auto-associative network captures the benchmark phenomenon. We then turn to the observational (Experiment 1) and unsupervised (Experiment 2) implementations, where the same architecture \emph{with no change in mechanisms} is shown to extend to learning problems where error-driven accounts struggle. 

\begin{figure}[!ht]
    \centering
    \centerline{\includegraphics[width=\linewidth]{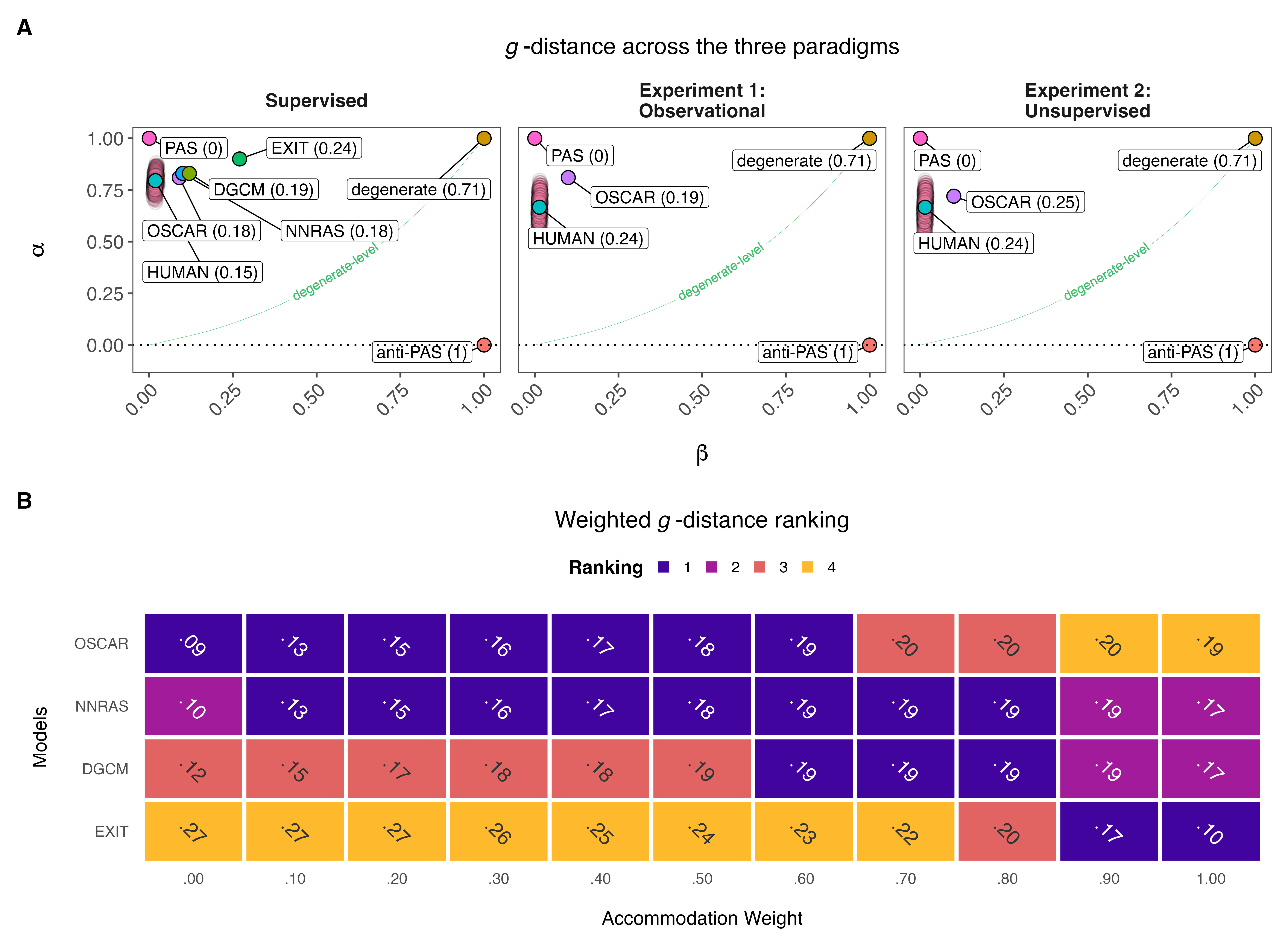}}
    \caption{\textbf{$g$-Distance across learning paradigms and $\alpha$-weighted sensitivity analysis of the model ranking in the supervised paradigm.} Figure (A) The three panels show models plus the split-half human performance (orange dots) in a two-dimensional space, where $\alpha$ (accommodation) and $\beta$ (breadth of unobserved model patterns) comprises this space. Colored dots correspond to models. Labels show the name of the model and their corresponding g. The degenerate line indicates random model performance. (B) The ranking of models according to different weighting of $\alpha$ and $\beta$. The y-axis shows the different models, whereas x-axis shows the different $w$ for $\alpha$. On the very left, models are purely evaluated on their ability to minimize unobserved model predictions, and on the very right, they are only evaluated on their ability to capture human ordinal patterns. OSCAR = Outcome-Specific Configuration-dependent Attention Representation; EXIT = EXemplar-based attention to distinctive InpuT; NNRAS = Neural Network with Rapid Attention Shifts; DGCM = Dissimilarity Generalized Context Model}
    \label{fig:gdistance}
\end{figure}

\begin{figure}[!hpb]
    \centering
    \centerline{\includegraphics[width=\linewidth]{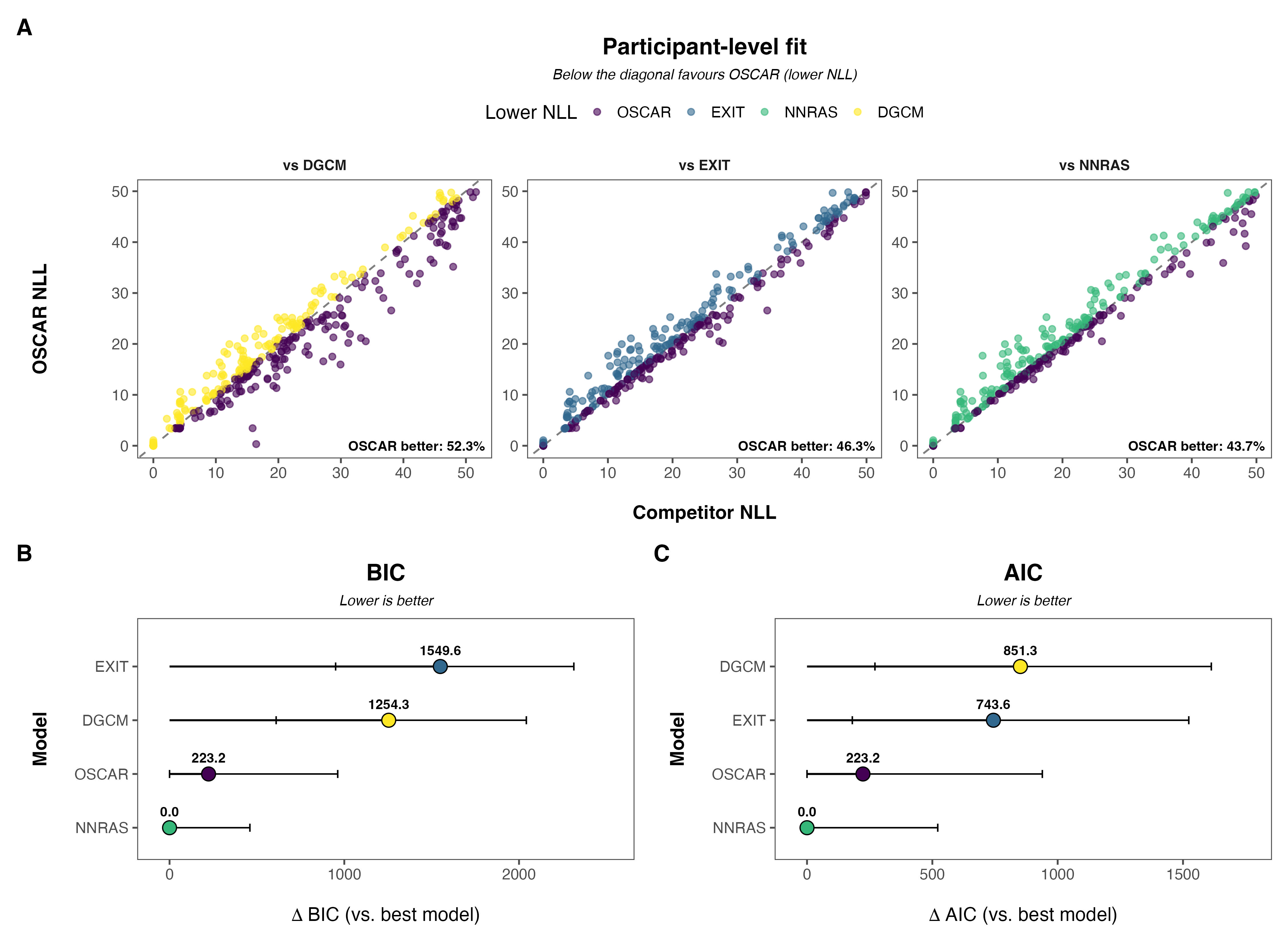}}
    \caption{\textbf{Model comparison on the supervised dataset.} (A) Participant-level negative log-likelihood (NLL) for OSCAR plotted against each competing model (EXIT, NNRAS, DGCM); points below the dashed identity line indicate participants for whom OSCAR achieved a lower (better) NLL, and point colour indicates which model achieved the lower NLL for that participant. The annotated percentage in each facet is the proportion of participants for whom OSCAR outperformed the corresponding competitor. (B) Difference in summed BIC relative to the best-fitting model ($\Delta$BIC; lower is better), with 95\% bootstrap confidence intervals (2,000 resamples of each model's per-participant BIC). (C) Difference in summed AIC relative to the best-fitting model ($\Delta$AIC; lower is better), with bootstrap confidence intervals as in (A). Across aggregate metrics (A, B, C), NNRAS provided the best fit overall, with OSCAR the closest competitor; at the participant level (C), OSCAR achieved lower NLL than DGCM for 52.3\% of participants but was outperformed by EXIT and NNRAS for the majority of participants (46.3\% and 43.7\% OSCAR wins, respectively).}
    \label{fig:fitting}
\end{figure}

Figure \ref{fig:gdistance}A shows the model performance on $g$ decomposed into $\alpha$ and $\beta$. The leftmost panel shows the model comparison between OSCAR and three competitors. OSCAR shares the win with NNRAS as the closest models to the PAS point with a  $g$ of $0.18$. This win comes through a reduction of $\beta$ relative to DGCM; OSCAR accommodates slightly fewer patterns than NNRAS but scores more on reduced flexibility relative to NNRAS. $g$ places DGCM in third place. EXIT remains the most able to capture the breadth of human ordinal patterns, but is also the one producing the most unobserved ones; consequently, $g$ places EXIT as the least adequate model. For this analysis, we have assumed an equal weighting for $\alpha$ and $\beta$. Figure \ref{fig:gdistance}B shows model rankings across different weighting of accommodation and breadth of unobserved model patterns. OSCAR and NNRAS are tied by $g$ across the majority of weightings ($0.10 \leq w \leq 0.50$) about the relative importance of $\alpha$ and $\beta$. OSCAR, NNRAS and DGCM are tied for $0.60$. NNRAS and DGCM take over under conditions of overwhelming disregard for flexibility ($0.70 \leq w \leq 0.85$); OSCAR is at the second place throughout this range. EXIT triumphs for the remaining range of $0.90 \leq w$, largely discounting model complexity. OSCAR and NNRAS remain the most adequate model for the widest range of $w$. These rankings converge with standard goodness-of-fit metrics, see Figure \ref{fig:fitting}, where the majority of metrics prefer NNRAS with OSCAR closely following. 

OSCAR is evaluated across all three procedures; the comparator models, by their reliance on a teaching signal, have no observational or unsupervised counterparts. For the observational and unsupervised evaluations, middle and rightmost panel of Figure \ref{fig:gdistance} respectively, OSCAR's performance on accommodation within the zone of model adequacy, and its flexibility remains comparable across the paradigms -- not increasing or decreasing substantially.

OSCAR produces the IBRE by combining outcome-specific attention with asymmetric weight structure: attention determines which input unit matters for a given possible outcome based on the feature configuration, while the excitatory and inhibitory connections encode the asymmetric task structure.

Attention concentrates on the shared feature A and on the opposing-outcome unit, and reverses across the common and rare output rows for the perfect predictors B and C -- attention to B is high when predicting rare (its absence is diagnostic) and near-zero when predicting common, with the converse for C. By the end of training, A is mutually excitatory with all units; B and C are mutually inhibitory with the opposing outcome and only weakly connected to their matched outcome unit (common and rare respectively). Because A is present on all trials, it forms strong excitatory connections to all units. The model learns equally strong inhibitory connections between B $\to$ rare and C $\to$ common, amplified through increased attention to those units, which counteract the excitatory connections formed between A and all other units, causing the model to learn to classify the training trial types correctly. The combination of these also underlie the characteristics of the inverse base-rate effect: base-rate following for A and the irrational response bias for BC. The model adjusts connection weights between A and all other units to accurately represent the frequency of correspondence between them. Attention to A is then learned to represent a similar pattern. For BC, the model learns to inhibit common output unit for the AC feature configuration more strongly, driven by the strong base-rate following learned for A, than to inhibit the rare output unit for AB feature configurations. This results in the inhibitory C $\dashv$ common link dominating the prediction, causing BC $\to$ common to be more strongly inhibited, and resulting in the BC $\to$ rare preference. The complete attention and weight matrices are presented in the Supplementary Figure \ref{fig:salience-weights}.

\begin{figure}[!htp]
    \centerline{\includegraphics[width=0.85\linewidth]{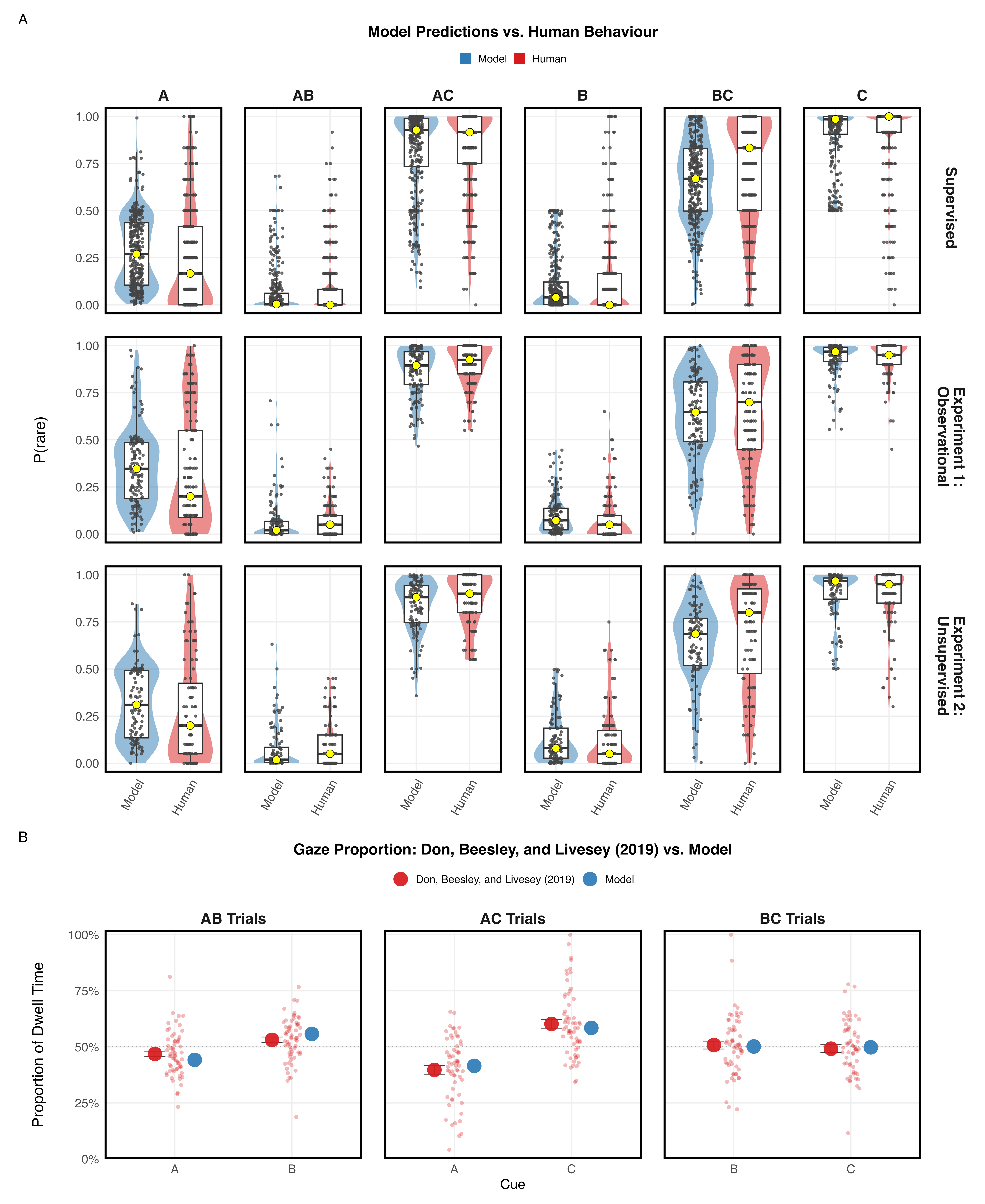}}
    \caption{\textbf{OSCAR reproduces the inverse base-rate effect across supervised, observational, and unsupervised learning}.  (A) Predicted P(rare) for each test cue (A, AB, AC, B, BC, C), comparing model (blue) to human (red) responses. Boxplots show medians and inter-quartile ranges, violins show full base-rate following for P(rare | A) together with the irrational response bias for high P(rare | BC) is reproduced in all three regimes. (B) Gaze dwell-time proportions by cue and trial type, human data (red, Don et al., 2019) vs. model (blue). Points: participant spread; large dots: means ± SEM; dotted line: chance.}
    \label{fig:simulations}
\end{figure}

Figure \ref{fig:simulations}A shows posterior predictive checks for each test items, where there is a good correspondence between OSCAR and human data, across all three paradigms. OSCAR reproduces all key qualitative patterns in the data across all paradigms.

\subsubsection{Mapping to fixation time}

This feature-specific fixation time reported by \cite{don2019learned} is a challenging result. During training, learners fixate more on B than on A on AB trials. Similarly, learners fixate more on C than on A on AC trials, and by a larger margin than B beats A on AB. This C $>$ B advantage is unique to training and does not persist into the test phase, where C $\approx$ B on BC trials. Previous models equated salience with visual attention, which was a reasonable assumption given that salience of a feature was a point-estimate.  Consequentially, models could account for eye-tracking data by acquiring higher salience for C relative to A on AC trials than for B relative to A on AB trials, all during training. But these point-estimates were not updated during the test phase, causing the model to produce more attention to C throughout the test phase, which contradicts to eye-tracking data showing comparable attention paid to both B and C on BC trials. 

By applying OSCAR to the group-level attention and weight matrices for the supervised condition from the last block of training, we replicate the full set of fixation tendencies observed by \cite{don2019learned}. For the simulations, we used the population\footnote{In differential evolutionary optimization algorithm, the population is the set of candidate parameter sets held at any one time; new candidates are generated from the differences between existing members, so the differences shrink automatically as the population converges as a function of the increasing number of iterations completed.} from the last iteration of our optimizer, centered on $\gamma = 0.58$. Figure \ref{fig:simulations}B shows the model predictions for the training items AC, AB, and the key test item BC. OSCAR  replicates the longer fixation for predictive features (B, C) over shared ones (A), including the C over B advantage during training. More importantly, OSCAR also recreates the B and C equivalence on BC trials, where the difference between B and C is negligible, something no other model is able to accommodate.

\section{Discussion}

The inverse base-rate effect (IBRE) is an irrational response bias characterized by an overestimation of rare events in the face of ambiguity. The mainstream theoretical account of this overestimation posits that error-driven attention drives this response bias. Across two experiments, we tested this central assumption.  

Experiment 1 implemented an observational IBRE: sentences paired symptoms with diseases, with no feedback or responses required--precluding explicit error. The IBRE still emerged. However, assumed symptom-disease causality may have invited implicit feature-to-label predictions. Experiment 2 removed this by using geometric shapes with no causal feature-label structure; the rare bias on BC trials persisted. Together, these results narrow the necessary IBRE conditions to two uniquely predictive features, a shared feature, sequential presentation, and the base rate, none of which carries a corrective error signal.

In order to explain these results, we developed OSCAR, an auto-associative feed-forward neural network that implements the core computational principles driving the IBRE: error-driven attentional learning. In this account, we proposed a (1) more granular representation of salience, where dimensional attention is represented with feature-specific vectors; and (2) equating features and labels within a shared input space. OSCAR is the only architecture that can produce the IBRE across all three paradigms. Furthermore, OSCAR is the only model that could account for eye-tracking data for both the training and test phase.

OSCAR explains IBRE via a combination of attention, inhibition, and excitation. This explanation is non-standard in the field. Most often, models explain the IBRE as a single salience value for C dominating responding on BC trials; an explanation that also fails to capture the ordinal B $\approx$ C fixation during test. Two departures from the standard account underlie this. The first is representational: attention is a matrix of outcome-specific dimensional vectors rather than the single per-feature salience assumed above. The second is architectural: the network is an auto-associator in which features and category labels share an input space, so the model reconstructs the complete pattern rather than predicting a label. The five consequences we discuss below follow from these theoretical commitments -- the first pair from the attention representation, the second from the auto-associative design -- and jointly they are what allow a single mechanism to produce the IBRE across the supervised, observational, and unsupervised procedures.

\subsection{Goal-directed attention}

The learning objective of attentional allocation is to minimize error, but the direction of allocation is based on what is being predicted on a given trial -- attention is not a fixed global property of a feature. In OSCAR's architecture, each row of the attention matrix is a specific predictive goal, and the excitability $\epsilon$ makes this concrete by gating which outcome units are live. The goal is therefore generated internally through a pattern completion mechanism rather than supplied as external feedback. Within each row, attentional values encode what features are relevant for a given outcome. The \cite{mackintosh1975theory} and \cite{kruschke2001toward} tradition attributes a single predictiveness-based salience signal for each feature. This single value collapses under multi-outcome learning because a single scalar cannot represent a more granular--diagnostic for rare, irrelevant for common--attentional map. A global point-estimate shared across multiple outcomes cannot hold this type of information \citep{dome2026shared}.

\subsection{Salience is modulated by feature configurations}

Attention to a feature depends on which other features are present. All competitive gating mechanisms implement this process in some form \citep{paskewitz2020dissecting,kruschke2001toward}, so configuration-dependence is not in itself distinctive. OSCAR implements a similar configuration-dependent attentional modulation, but its driving mechanism operates over a matrix representation that enables the same feature to receive different attention across compounds. This is what recovers the fixation dissociation reported by \cite{don2019learned}, where the C-over-B advantage present during training does not persist to BC at test (where the two are fixated comparably) -- a pattern point-estimate gating cannot produce. The product-of-expert mapping, as we formalized it here, is the driving force that takes this more granular representation and maps it to predictions about fixation patterns. Within this machinery, we specified attention as uncertainty-averse \citep{speekenbrink2022chasing,stojic2020its}, which is also consistent with the tradition of \cite{mackintosh1975theory}.

\subsection{Feature absence is diagnostic}

An auto-associator is designed to reconstruct the whole pattern (feature space) including absent dimensions.  For example, B encodes both the presence of A (excitatory link) and the absence of C and rare (inhibitory links), which results in B acquiring higher attention for those connections. More generally, features simultaneously carry information about the presence or absence of other features. Feed-forward outcome predictors cannot do this, because they predict labels and not the complete pattern. As a consequence, OSCAR can represent directionally opposing information features carry about the presence or absence of other features.

\subsection{Outcomes acquire salience}

Because the architecture is auto-associative $m = n$, outcomes are encoded in the same feature space as inputs; the teaching vector $t = \epsilon + s$ combines the present features and the correct outcome within that single features space. Outcome dimensions are themselves attendable and accrue salience. In the unsupervised implementation, the labels are features, so outcomes stop being a privileged category and are subject to the same attentional shift and weight update processes. This is what allows the unsupervised case to follow from the same machinery rather than requiring a separate account, satisfying the third of our theoretical requirements.

\subsection{Implications of uncertainty measures for confidence}

The model calculates two measures of uncertainty from learned representations, which it maps to eye-tracking via a feature-specific score, $F$. We propose two further behavioral mappings for these estimates: confidence and reaction times. Confidence is one possible operationalization of metacognitive monitoring \citep{flavell1976metacognitive,flavell1979metacognition,fleming2024metacognition}, a process that introspects one's own certainty. Confidence and response time are reliably negatively correlated, with higher confidence associated with faster responses, a relationship documented for over a century \citep{henmon1911,volkmann1934,vickers1979,ratcliff2013,weidemann2016}. \citet{kiani2014choice} further show that choice certainty is informed jointly by the state of the evidence and by elapsed decision time, tying the two measures to a common graded quantity derived from the evidence. Extrapolating our modeling framework, OSCAR predicts that $dec(S)$, one component in the feature-specific fixation scores, $F$, modulates reaction times and confidence reports in addition to fixation times. The variable $dec(S)$ indexes the degree of consensus among features and is therefore a latent source of response competition. Based on these, OSCAR predicts that the low-consensus on BC trials corresponding with higher-uncertainty will produce longer reaction times, lower confidence relative to low-uncertainty test trials such as AC and AB.

\subsection{Limits}

There are neural recordings that show C to be more dominant than B when presented by themselves during test. \cite{inkster2022neural} showed that number of brain areas associated with prediction error are more active for C than B during the test phase. Similarly, \cite{wills2014attention} showed that event-related potential also shows C $>$ B preference, in an attention-like way. At the moment, it is not clear how OSCAR would produce this C $>$ B difference only from attention matrices. Connection weights show that C $\dashv$ common is stronger than B $\dashv$ rare, which could give some indication about where this directional relationship can be identified within OSCAR, but the exact mapping between them and the neural correlates of IBRE is unclear.

\section{Conclusion}

Across two experiments, we demonstrated that the IBRE does not require a corrective teaching signal. Existing models cannot accommodate this. Error-driven attention models optimize a supervised objective whose targets our procedures do not provide, and test-phase-only accounts cannot represent the training manipulation at all. 

OSCAR removes the dependence on supervision within an auto-associative architecture: features and category labels share a single pool of units, and learning is auto-associative pattern completion over that pool, so the target is the activation pattern itself and the error signal is endogenous. The IBRE emerges through a cooperation of outcome-specific competitive attentional gating and a group of excitatory and inhibitory connections. The same architecture and the same objective carry over without modification to the observational and unsupervised procedures. OSCAR is the only model to produce the IBRE in all three; and it alone reproduces the distributed pattern of visual fixation across both training and test -- a chimera of results no previous model could accommodate.

This locates the effect in the architecture. The rare bias is an errorless irrational tendency: it requires no supervised error signal, only an endogenous self-generated signal whose target the network supplies for itself through pattern completion. The irrationality is an emergent property of the architecture that has learned the regularities of its environment -- not the residue of optimizing against an external teacher.

\section{Open Science}

We have made available the two experiments written in javascript, the analysis code, the raw data, and all other supplementary materials both on the Open Science Framework and GitHub. Experiment 1 is shared on \href{https://osf.io/auwvt/}{https://osf.io/auwvt/}, and \href{https://github.com/lenarddome/ply216-observational-ibre}{https://github.com/lenarddome/ply216-observational-ibre}. Experiment 2 is similarly shared on \href{https://osf.io/2tmc4/}{https://osf.io/2tmc4/} and \href{https://github.com/lenarddome/ply222-non-causal-ibreBA}{https://github.com/lenarddome/ply222-non-causal-ibre}. All simulation code can be found on \href{https://github.com/lenarddome/tue008-full-network-model}{https://github.com/lenarddome/tue008-full-network-model}.

\section{Acknowledgement}

We thank Selena Lockett and Charlotte Hubbard for the help and contributions in the data collection for Experiment 1. A preliminary partial report of Experiments 1 and 2 was published in the Proceedings of the 45th Annual Meeting of the Cognitive Science Society \citep{dome2023errorless}.

\bibliographystyle{abbrvnat}
\bibliography{references}

\newpage
\appendix

\section{Parameter Bounds}
\label{section:psp}

The parameter bounds used to define the parameter space and the search space for the differential evolutionary optimizer are presented on Table \ref{tab:adaptive-bounds}.

\begin{table}[!ht]
  \caption{Lower and upper bounds of model parameters.}
  \label{tab:adaptive-bounds}
  \centering
  \begin{tabular}{rccccccc}
      \toprule
      Model &  $\alpha$ & $\phi$ & $\mu$ & $\rho$ & $P$ & $c$ & $\sigma$\\
      \hline
      \\
      OSCAR & $[0, 1]$ & $[0, 10]$ & $[0, 1]$ & $[0, 10]$ & $[1, 10]$ & & \\
      NNRAS & $[0, 1]$ & $[0, 10]$ & $[0, 1]$ & $[0, 10]$ & $[1, 10]$ & & \\
      EXIT  & $[0, 1]$ & $[0, 10]$ & $[0, 10]$ &  $[0, 10]$ & $[1, 10]$ & $[0, 1]$ & $[0, 1]$ \\
      \\
      \hline
      Model & $w_k$ & $s$ & $c$ & $\beta_{A}$ & & & \\
      \hline
      \\
      DGCM & $[0, 1]$ & $[0, 1]$ & $[0, 10]$ & $[0, 1]$ & \\
      \\
      \bottomrule
  \end{tabular}
\end{table}

\section{Goodness-of-fit metrics}

No single goodness-of-fit metric is independent of the assumptions under it is computed. Here, we have decided to report multiple metrics that could be calculated from negative log-likelihood on two grounds: (1) different metrics characterize different degrees of fit-complexity tradeoff, and (2) convergence across metrics with different assumptions provides a robustness test against the specific failure mode of either; where metrics diverge, the divergence itself is diagnostic of under what assumption the model is being rewarded.

\begin{table}[!ht]
\caption{Definitions of model fit and selection metrics.
  $k$ = number of free parameters; $n$ = observations per participant;
  $N$ = number of participants; $J$ = response alternatives (2).}
\label{table:metrics}
\centering
\begin{tabular}{@{}llp{7cm}@{}}
\toprule
Metric & Formula & Description \\
\midrule
$\Sigma\text{NLL}$ &
  $\displaystyle\sum_i \text{NLL}_i$ &
  Sum of per-participant negative log-likelihoods at best-fit parameters; lower is better. \\[8pt]
$\Sigma\text{BIC}$ &
  $\displaystyle\sum_i \!\left(\text{NLL}_i + \tfrac{1}{2}k\ln n\right)$ &
  NLL penalised by model complexity measured as a number of parameters scaled by log sample size; asymptotically consistent model selector; lower is better. \\[8pt]
$\Sigma\text{AIC}$ &
  $\displaystyle\sum_i \!\left(\text{NLL}_i + k\right)$ &
  NLL with a fixed per-parameter penalty targeting predictive accuracy rather than model identification; lower is better. \\[8pt]
\bottomrule
\end{tabular}
\end{table}

\section{Model weights and saliences}

\begin{figure}[!ht]
    \centering
    \includegraphics[width=\linewidth]{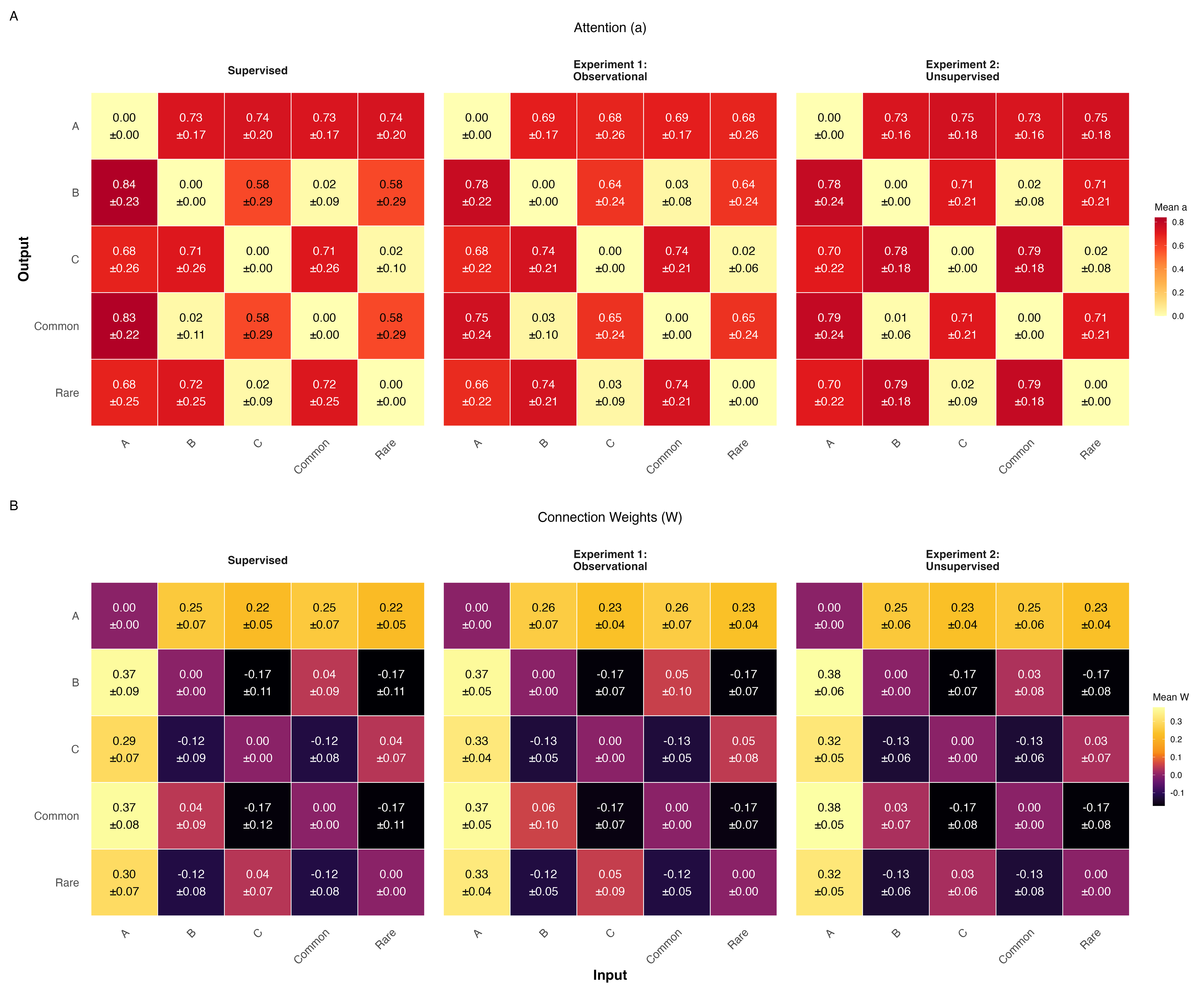}
    \caption{(A) Mean outcome-specific attention weights $a_{ij}$ from each input unit (columns within a panel) to each output unit (rows), averaged across simulated participants ($\pm$ 1 SD); the diagonal is zero by architectural constraint. (B) Learned connection weights $W_{ij}$, same layout.}
    \label{fig:salience-weights}
\end{figure}

Below we present the connection and attention weights by the end of training across all three paradigms in Figure \ref{fig:salience-weights}. All are taken from subject-level runs at the corresponding best-fitting parameter values; they are intended to make the learned microstructure inspectable, not to establish that this particular structure is unique to those values. In all paradigms, OSCAR explains the inverse base-rate effect the same way. The connection weight matrices give the auto-associative mapping acquired between feature and outcome units; the attention matrices give the outcome-specific gains applied to that mapping during decision and later updates.

\end{document}